\documentclass[11pt]{article}

\usepackage[hidelinks]{hyperref}
\usepackage{fullpage}
\usepackage{url}
\usepackage{graphicx}
\usepackage{caption}
\usepackage{subcaption}
\usepackage{amsmath}

\begin{document}
\title{Checkpointing as a Service in Heterogeneous Cloud Environments}

\newcommand*\samethanks[1][\value{footnote}]{\footnotemark[#1]}

\author{
\makebox[0.8\textwidth]{Jiajun Cao$^\dag$\thanks{This work was partially
  supported by the National Science Foundation under Grants ACI-1440788
  and OCI~1229059, and by a grant from Intel Corporation.} \hfill 
Matthieu Simonin$^\ddag$  \hfill
Gene Cooperman$^\dag$\samethanks{$^*$}  \hfill
Christine Morin$^\ddag$}  \bigskip \\
$^\dag$College of Computer and Information Science \\
  Northeastern U., Boston, MA, USA \\
  \{jiajun,gene\}@ccs.neu.edu
\medskip
\and
$^\ddag$Inria \\
  IRISA, Rennes, France\\
  Matthieu.Simonin@inria.fr \\
  Christine.Morin@inria.fr
}
\date{}

\maketitle

\begin{abstract}
A non-invasive, cloud-agnostic approach is demonstrated for extending
existing cloud platforms to include checkpoint-restart capability.
Most cloud platforms currently rely on each application to provide
its own fault tolerance.  A uniform mechanism within the cloud itself
serves two purposes:
(a)~direct support for long-running jobs, which would otherwise
require a custom fault-tolerant mechanism for each application; and
(b)~the administrative capability to manage an over-subscribed cloud
by temporarily swapping out jobs when higher priority jobs arrive.
An advantage of this uniform approach is that it also supports parallel
and distributed computations, over both TCP and InfiniBand, thus allowing
traditional HPC applications to take advantage of
an existing cloud infrastructure.
Additionally, an integrated health-monitoring mechanism detects when
long-running jobs either fail or incur exceptionally low performance,
perhaps due to resource starvation, and proactively suspends the job.
The cloud-agnostic feature is demonstrated by applying the implementation
to two very different cloud platforms: Snooze and OpenStack.  The use
of a cloud-agnostic architecture also enables, for the first time,
migration of applications from one cloud platform to another.
\end{abstract}

\section{Introduction}
Cloud computing provides users with the illusion of an infinite
pool of resources available over the Internet, from which they can
access on~demand and through self-service the resources they need for
their applications. In less than a decade numerous cloud providers
have flourished, each of them operating one or several data centers
in different locations. Cloud providers target transparent failure
and maintenance management, with the twin goals of satisfying their
customers, and providing the high resource utilization that maximizes
their profit. Many failure, reconfiguration and resource management
strategies rely on the ability to migrate virtual machines (VMs)
both between data centers and within a single data center.  Customers
want their applications to be executed reliably in the cloud, and
they seek to escape the vendor lock-in phenomenon by taking advantage
of a market of heterogeneous clouds.

We propose a novel \textit{Checkpointing as a Service} approach, which enables application checkpointing and migration in heterogeneous cloud environments. 
Our approach is based on  a non-invasive mechanism  to add fault tolerance to an existing
cloud platform {\em after the fact}, with little or no modification
to the cloud platform itself. It achieves its cloud-agnostic property
by using an external checkpointing package,
independent of the target cloud platform.

Such a cloud-agnostic checkpointing service is important for at
least three distinct reasons:
\begin{enumerate}
  \item provision of fault tolerance for long-running tasks;
  \item improved cloud efficiency (low-priority applications can be
	suspended to stable storage, and restored only when idle CPU
	cycles are available); and
  \item migration of tasks between distinct IaaS clouds (e.g., between
	one operated by  the Snooze system and one operated by the OpenStack system).
\end{enumerate}

The proposed  {\em Cloud-Agnostic Checkpointing Service} (CACS) is 
retro-fitted into multiple cloud platforms.   This is demonstrated for
two cloud platforms:  Snooze and OpenStack.

A necessary component of CACS is a health-monitoring mechanism
that not only detects when an application has ``died'', but generally
when an application is unhealthy.  Detecting the latter is non-trivial,
since only the application developer knows if the termination or
non-responsiveness of one
process is fatal to the overall computation.  Hence, a hook is provided
for each application to determine its own ``health''.

Our contributions are three-fold:
\begin{itemize}
  \item We provide the first transparent checkpointing scheme for
	centralized, parallel and distributed computations in the Cloud.
  \item The transparent checkpointing scheme is {\em cloud-agnostic}.
	The minimal assumptions of this approach allow it to
	be extended to most cloud architectures.
  \item Migration of computations among heterogeneous clouds is provided.
\end{itemize}

CACS employs the DMTCP
package~\cite{AnselEtAl09}, a checkpointer for distributed multithreaded
applications.  This was chosen for its transparent support of
distributed applications, including both TCP/IP and the InfiniBand
network~\cite{CaoEtAl14}.

Moreover, we show that our approach toward
checkpointing and migration scales with application size and with the
number of applications hosted in a data center. 

The remainder of this paper is organized as follows.
Section~\ref{sec:motivation} presents further background and motivation
for the approach.
In Section~\ref{sec:methodology} we discuss  the principles that guided the design of the cloud checkpointing service.
Section~\ref{sec:implementation} provides an overview of CACS. Section \ref{sec:scenarios} presents some typical scenarios,
from application submission through checkpoint, recovery and/or migration
to a new cloud, and finally application termination.
Section~\ref{sec:prototype} presents our prototype implementation.
In Section~\ref{sec:experimentalEvaluation} we analyze the results
from a first experimental evaluation. Section~\ref{sec:relatedWork}
describes the related work, while the conclusions are presented in
Section~\ref{sec:conclusion}.

\section{Motivation}
\label{sec:motivation}

The cloud-agnostic checkpointing service is intended to provide a
single checkpointing solution for heterogeneous computing services.
This eliminates the need for each computing service to implement its
own checkpointing solution.  Such solutions are required not only for
long-lived computations, but also for numerous other use cases.

\subsection{Context:  IaaS and Heterogeneity among Clouds}
\label{sec:context}

IaaS clouds provide resources on demand to their customers in the form of
virtual machines. IaaS clouds are heterogeneous, each coming with
its own marketplace of {\em Virtual Machine Images} (VMI). Customers of an IaaS
cloud provider select VMIs among those offered. Different VMIs correspond
to different, possibly customized, combinations of an operating system
kernel, an OS~distribution, and a processor architectures (32- or 64-bit).
Instances are characterized by the amount and type of resources they use
(e.g., number of cores, memory capacity, disk capacity).

The IaaS cloud management system manages the life cycle of a VM 
from submission to termination. In particular, it allocates the server resources 
among the VMs and performs VM scheduling. Servers and VMs are monitored
in order to determine efficient resource management strategies.  In a
cloud environment, a distributed application is executed using a set of
interconnected VMs called a {\em virtual cluster}.

Different cloud environments also introduce heterogeneity among dimensions
other than those described in the previous paragraph.  Servers may have
different hardware configurations (e.g., Intel versus AMD processors),
and may run different combinations of VMs and hypervisors
(e.g., KVM/QEMU, Xen, and Linux containers).

A sufficiently robust checkpoint-restart package, such as the DMTCP
package used here, can overcome these sources of heterogeneity.  As a
prerequisite, an end user must design her applications for a common
denominator:  compiling for the intersection of Intel and AMD instruction
sets, avoiding the most recent system calls in order to provide backward
compatibility, programming scalability to adjust for fewer or more cores,
and even compiling for a 32-bit instruction set if a combination of
32- and 64-bit CPUs is anticipated.  In this way, a cloud-agnostic
checkpointing service can directly migrate applications among such
heterogeneous resources.

Finally, a cloud-agnostic checkpointing service must be tolerant
of the different types of IaaS cloud management systems that exist today.
These include OpenStack~\cite{OpenStack} (widely adopted in production data centers),
Nimbus~\cite{keahey2008science} (targeting  scientific computing), and Eucalyptus~\cite{nurmi2009eucalyptus},
OpenNebula~\cite{milojivcic2011opennebula} and Snooze~\cite{Feller:2012:SSA:2310096.2310230}
(originating from academia), and Amazon~EC2 (the most widely used public commercial cloud). 
 IaaS cloud systems may use different VM disk image 
formats (e.g. QCOW, VMDK) and provisioning methods. They
may provide different APIs for storage and VM management. However, some
popular interfaces (EC2 for VM management and S3 for storage, for example)
have become de-facto standards. Recently, there have been a number of
emerging standard APIs such as DMTF CIMI ~\cite{cimi} and OGF OCCI~\cite{occi},
which have not yet become mainstream.

\subsection{Use Cases}
Many motivating use cases demonstrate the need for a portable
efficient cloud checkpointing service.  A first use case is
fault-tolerant application execution in the cloud. Long-running jobs
(such as OpenMP-based or MPI-based scientific applications) should be
periodically checkpointed, so that they can be restored from their last
checkpoint in the event of a failure.

Ideally, it should be possible to restore an application either in
the same data center or in another one from the same cloud provider to
survive catastrophic failures affecting a whole data center.  However,
although a second data center may be available, it may be running under a
different type of infrastructure.  This gives rise to the second use case:
migration among heterogeneous clouds.

A third use case occurs when the cloud provider needs to transparently
carry out maintenance operations.  Providers can stop all applications
and checkpoint them or migrate them to other clusters, before taking
down a cluster for maintenance.

A fourth use case occurs in the scientific world, in the framework of
advanced VM scheduling algorithms.  Periods of low demand may lead to
potentially low utilization rates.  A VM scheduler attempts to increase
resource utilization.  Opportunistic preemptible leases running on
backfill VMs have been proposed for this case by Marshall
\hbox{et al.}~\cite{5948611}. Such leases give a user access to a resource
at an indeterminate time and for an indeterminate amount of time, but are less
expensive than traditional on-demand leases.  Transparent cloud-agnostic
checkpointing allows any  scientific application to use such a lease.

Proactive cloud migration provides a fifth use case.  For a cloud
provider operating multiple possibly heterogeneous data centers it is
desirable to be able to migrate VMs from
one cloud to another.  Energy-efficient resource management
policies such as follow-the-sun  (aimed at exploiting renewable energy
sources to the extent possible) and cloud bursting are two illustrating use cases~\cite{Wood:2011:CDP:2007477.1952699} .

A sixth use case is vendor lock-in.  Cloud customers currently face
vendor lock-in issues in re-targeting their distributed applications
from one cloud provider to another.  A cloud-agnostic checkpointing
service would overcome heterogeneity issues and empower cloud users to
take advantage of the competitive cloud computing market by outsourcing
their applications to another provider.

Last but not least is the seventh use case.  Migrating legacy
distributed applications to the cloud remains a tedious task for users who
don't have system administration skills. In the context of IaaS clouds,
porting from a cluster to a virtual cluster in the cloud may require the
skills of a system administrator and the domain-specific knowledge
of an end-user. The portable checkpointing service proposed here
is a key building block for a \textit{cloudification} service,
significantly reducing the burden of legacy applications users in moving
their application to the cloud. In principle, a user would simply use
the CACS-based cloudification service to migrate her application from
her desktop or local cluster to a selected IaaS provider, since the design
of CACS will extend to run on other
resource management services, including a Linux desktop system and the
resource management system (RMS) (e.g., batch system) of an HPC cluster.

There is no claim that the current CACS design will satisfy all of the
above use cases. Some use cases might require specialized cloud configurations
or specialized data services~\cite{ghoshal2012frieda}.

\section{Design Principles of CACS}
\label{sec:methodology}

To address the requirements presented in the previous section, we
developed a Cloud-Agnostic Checkpointing Service. We discuss five
principles that guided its design.

\subsection{Why Using a Process-level Checkpointer rather than VM Snapshots?}
A key design principle of the Cloud-Agnostic Checkpointing Service
is that it leverages a process-level checkpointer
for checkpointing distributed applications executed in virtual
machines. There are two primary reasons why a process-level checkpointer
was chosen instead of
using the VM snapshot mechanism offered by hypervisors.
First, snapshotting a set of virtual machines is more expensive than
checkpointing a set of processes. In the latter case the operating
system is not checkpointed, and the checkpoint size is much smaller. While  data deduplication techniques~\cite{Wood:2011:CDP:2007477.1952699} can be used to reduce the cost of live migration, our approach 
has a broader applicability being hypervisor-agnostic. 
Second, process-level checkpointers like DMTCP manage
dependencies among communicating multithreaded processes when saving
a checkpoint. When checkpointing a distributed application running in
multiple VMs using VM snapshots, hypervisors fail to handle the
inter-process communications of distributed processes.

VM snapshots have been extensively used to checkpoint an application
running in a single VM, since it provides a generic checkpointing mechanism
transparent to the application, which does not need to be modified.
A process-level checkpointer like DMTCP is fully transparent
to the application and generic, including support for checkpointing sets of
communicating multi-threaded processes.  Moreover, in an environment
of multiple heterogeneous clouds,
a process-level approach to checkpointing the
distributed applications of a virtual cluster provides better
portability and interoperability than one
based on VM snapshots.  This avoids the difficulty of porting VM images and
adapting to multiple IaaS cloud management APIs, when dealing with
different cloud management systems.

\subsection{Eliminating the Checkpoint Management Burden}

Checkpointing should come as a service, implying a minimal burden for
users. In our approach, users  request their VMs from CACS rather than
directly  from the IaaS cloud manager, and submit their application to CACS
while specifying the checkpointing policy (e.g., checkpoint frequency). 
CACS obtains the VMs, installs and configures the process-level
checkpointer and the application inside the VMs, and then automatically
triggers checkpoints according to the user-defined policy.

\subsection{Portability and Interoperability}
CACS has been designed to execute on top of unmodified
existing IaaS cloud management systems, to address a broad IaaS cloud
market.  Thus, it relies on the de~facto standard APIs offered by
most IaaS clouds systems, namely EC2 for VM and S3 for storage
management.

An important requirement is to be able to detect failures at the
level of the server, the VM and the application, within the underlying
IaaS cloud management system.  For instance, OpenStack does not
provide an API to report infrastructure failures to clients. So
CACS must include a cloud-agnostic monitoring system. Yet at the
same time, CACS should be able to exploit any existing monitoring
mechanisms of the underlying IaaS cloud where they exist, as in the
case of the Snooze VM management system.

Another portability issue arises from the fact that different IaaS
management systems may use different VMI formats and offer different
types of VM.  This further motivated the first design decision: to
use application-level checkpointing rather than VM snapshots.

\subsection{Scalability}
CACS should scale with the number
of concurrent VMs so that it can be used to tolerate failures in data
centers; and it should scale with the size of the applications (with the number of
VMs per application) so as to have a limited impact on the execution time of
large distributed applications.

The choice of implementation for stable storage has an important impact
on these two types of scalability.  Thus, CACS relies on
distributed parallel file systems such as Ceph~\cite{weil2006ceph} in order to cope with
the huge volume of data to be stored when several checkpoints are taken
simultaneously.
Similarly, efficient VM management is also essential to limit as much
as possible the recovery time.

\subsection{Usability}

Nowadays, providing a REST API for a service is a key feature. Resources are served using various server
representations.  This eases the interaction with users and with third-party software (e.g., CLI, web-based GUI).
Moreover, the statelessness of server requests in a REST API provides
for server scalability, since communication between clients and server
is loosely coupled.

\section{Overview of CACS}
\label{sec:implementation}

Next, the CACS architecture and its core components are presented. First,
the underlying technology, DMTCP, is introduced. Then the CACS internal
components are examined.

\subsection{DMTCP Application Checkpointer}

\label{sec:dmtcp}

The choice of DMTCP (Distributed MultiThreaded
CheckPointing)~\cite{AnselEtAl09} for checkpoint-restart was dictated
by the maturity of that ten-year old project~\cite{CoopermanAnselMa05}
available as binary
packages for Debian/Ubuntu, Fedora/CentOS/Scientific~Linux/Red~Hat,
and OpenSUSE).  In particular, DMTCP supports the types of
migration of processes among heterogeneous environments that were
described in Section~\ref{sec:motivation}.

In DMTCP, each  application is associated with
a unique DMTCP application coordinator in charge of coordinating the checkpointing
of processes running on distinct computer nodes.   The coordinator need not reside
on a host that is hosting application processes, and directly communicates with DMTCP daemons running on each node hosting application processes.
When an application is restarted, a new DMTCP
coordinator is used, thus avoiding any single point of failure.

\subsection{Architecture of CACS}

\begin{figure}[ht!]
\centering
\includegraphics[width=0.5\columnwidth]{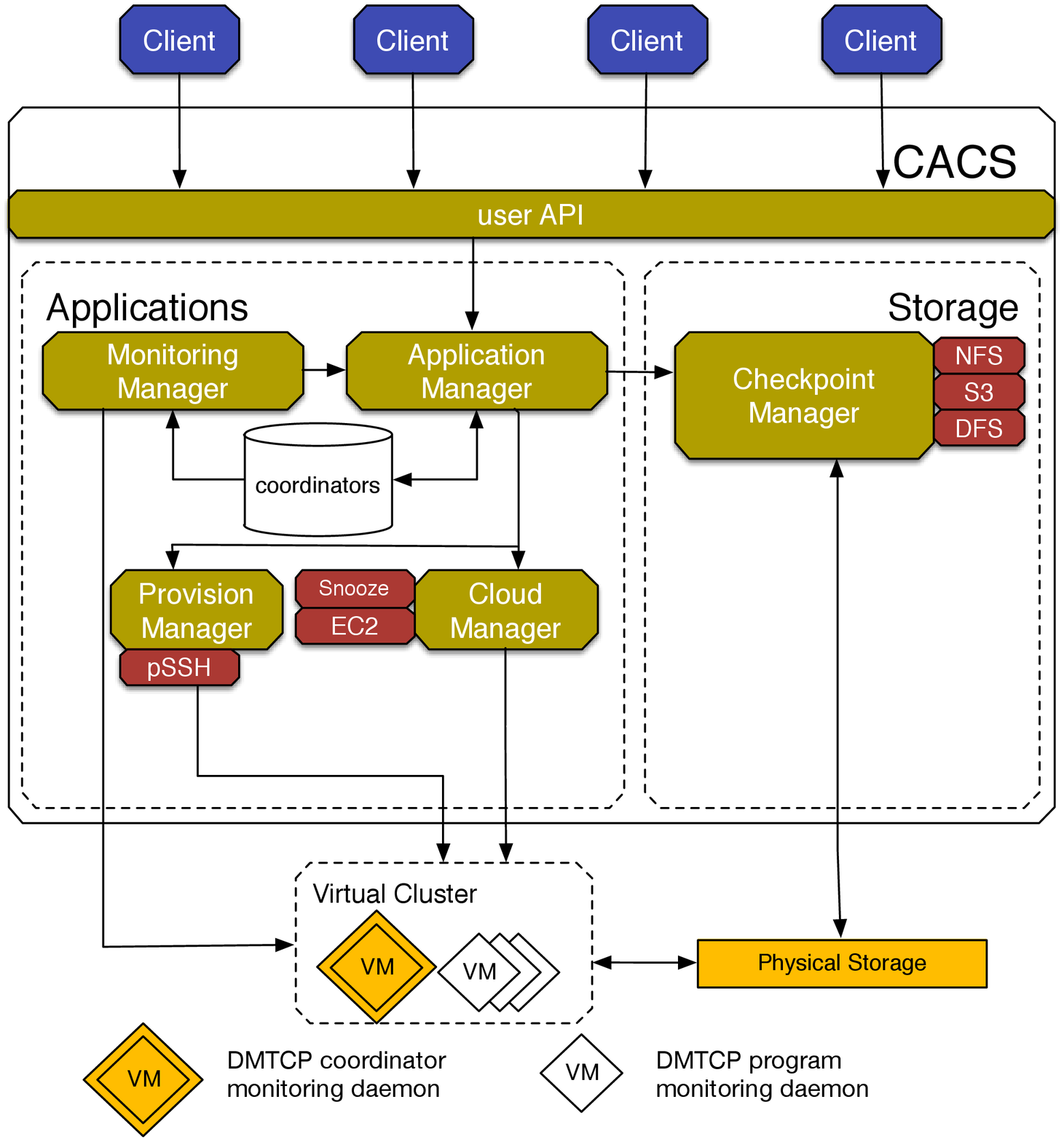}
\caption{\label{fig:arch} Cloud Checkpointing Service Architecture}
\end{figure}

Figure \ref{fig:arch} depicts the overall architecture of the service.
The \emph{RESTful API} allows users to manage their applications and their corresponding checkpoints.
The \emph{Coordinators Database}
stores all the  applications information.
The {\em Application Manager} orchestrates application management (start and restart) and enforces failure recovery mechanisms. 
It communicates with the {\em Cloud Manager} to manage (create, destroy) virtual clusters. The  Cloud Manager interacts with the underlying IaaS cloud system to manage the VMs.
The {\em Provision Manager} takes on the burden of efficiently configuring the virtual clusters.
The {\em Checkpoint Manager} component is for managing
the application checkpoint images. 
The {\em Monitoring Manager} component enables VM and application process failure detection. It is notified about application health issues and VM failures by monitoring daemons running in each VM of the virtual cluster executing the application. In the event of a notification, it interacts with the Application Manager in order to stop and/or recover the application from a previous checkpoint.

An application is executed under the control of DMTCP whose  daemons run in each VM hosting the application processes, with one of them running the DMTCP coordinator.

\section{Typical Scenarios}
\label{sec:scenarios}

We describe five scenarios of a typical CACS user, ordered according to
the life cycle of an application comprising \textit{n} processes running in  \textit{n} different virtual machines.
This section describes in greater details some mechanisms used to handle user requests. Table~\ref{table:api} depicts the description of the resources managed by the API and the available operations.

{\scriptsize
\begin{table}[ht!]
  \begin{tabular}{p{0.45\columnwidth} p{0.55\columnwidth}}
    \hline
    {\tt coordinators resource } \\
    GET /coordinators & returns the list of coordinators known by the system \\
    POST /coordinators & add a new coordinator to the system \\
    \hline
    { \tt coordinator resource } \\
    GET /coordinators/:id & returns the information of the coordinator with
		 id : id \\
    DELETE /coordinators/:id & delete the coordinator \\
    \hline
    { \tt checkpoints resource } \\
    GET /coordinators/:id/checkpoints & returns the list of the checkpoints
		of the coordinator \\
    POST /coordinators/:id/checkpoints & trigger a checkpoint for the
		coordinator or upload a checkpoint image \\
    \hline
    { \tt checkpoint resource }\\
    GET /coordinators/:id/checkpoints/:id & returns the information
		 of a checkpoint \\
    POST /coordinators/:id/checkpoints/:id & restart the coordinator
		 from the checkpoint \\
    DELETE /coordinators/:id/checkpoints/:id & delete the checkpoint \\
  \end{tabular}
  \caption{REST API description}
  \label{table:api}
\end{table}
}

\subsection{Application Submission}
\label{sec:scenarios:submission}

Here We describe application submission to CACS.
A POST request is issued to the {\tt coordinators resource} and
the body contains the representation of an { \em Application
Submission Request (ASR)}.  The ASR encapsulates the VM templates
and the configuration parameters of DMTCP needed
to start the application.

\begin{figure}[ht!]
\centering
\includegraphics[width=0.5\columnwidth]{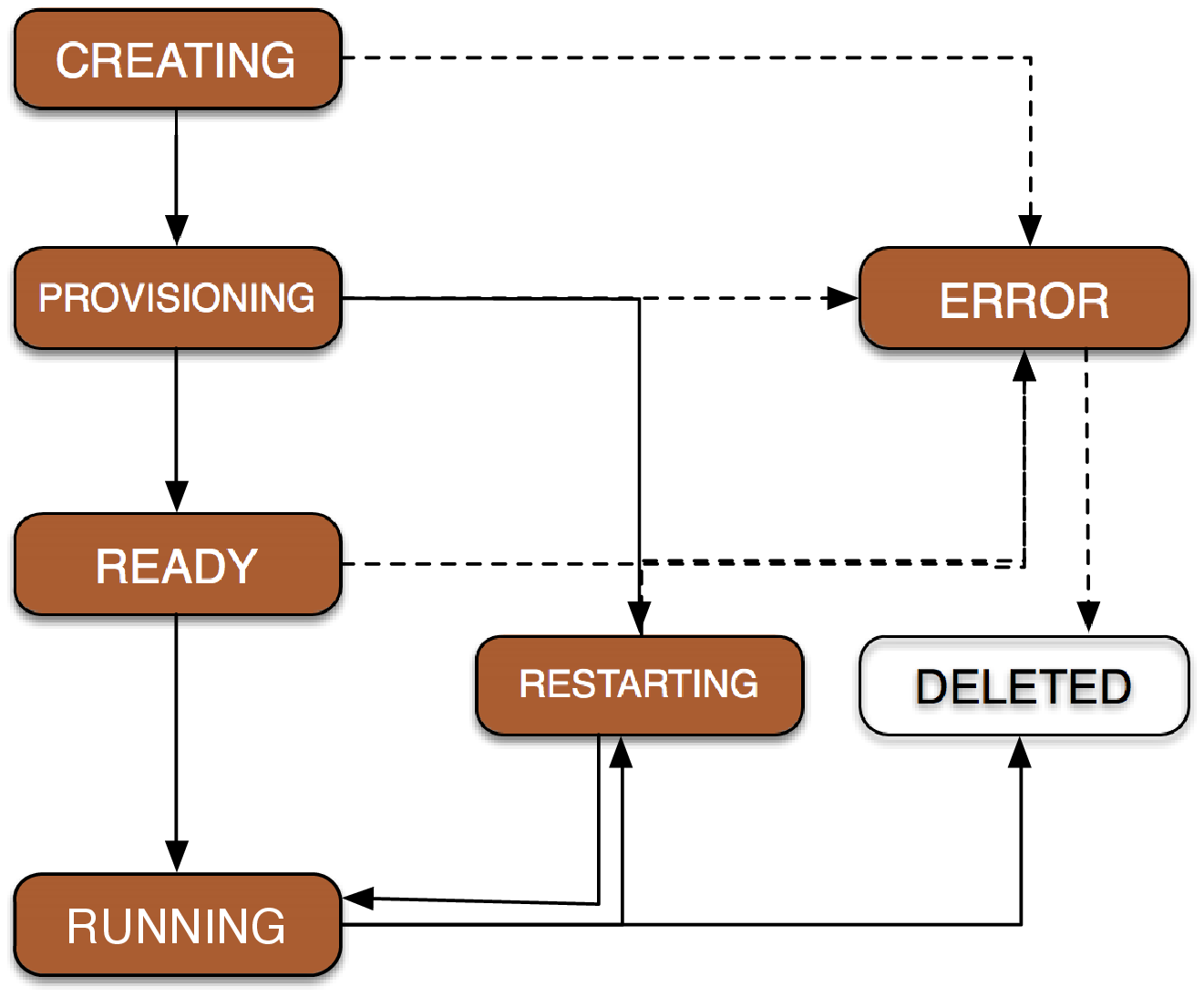}
\caption{\label{fig:states} CACS Coordinator states }
\end{figure}

Once the Application Manager validates the ASR, the application enters the CREATING 
phase (see Figure \ref{fig:states}) during which virtual resources are claimed by the Cloud Manager. 
Once the VMs have been given to the computation, the
PROVISION phase starts. In this phase, the Provision Manager remotely executes  specific
commands to prepare the computation to be run.
The provision includes internal actions
 (e.g., creation of checkpoint directory in the VMs)
but also user-defined configuration (e.g., periodicity of the checkpoints,
specific initializations).  The provisioning phase may differ according
to the storage back-end used.

The READY state is introduced to reflect the fact that all the VMs are
ready to start the computation.  The RUNNING state indicates that the
computation is in progress. In this phase, checkpoints can be saved.

An alternative way of starting an application is described in section
\ref{sec:scenarios:migration}.

\subsection{Saving Checkpoints}
Three modes of transparent checkpointing are supported:
(1)~user-initiated checkpointing;
(2)~periodic checkpointing; and
(3)~application-initiated checkpointing (for example,
	at the end of each application iteration).
The first case can be fulfilled by issuing a POST request
to the corresponding {\tt checkpoints resource}. In the 
second and third case, DMTCP triggers the checkpoint without
the need for a POST request. CACS distinguishes between local
and remote storage. Where fast local storage is available (e.g., a local
disk, an SSD, or a RAM~disk inside RAM itself), the checkpoint image
is written first to the local storage.  For redundancy, it is also
copied (on a lazy basis) to remote storage, such as Ceph and NFS.

\subsection{Application Recovery, Cloning and Migration}
\label{sec:scenarios:migration}
The API enables the following scenarios:
(1)~application restarting (the application state is reset to a
  previous checkpointed state and restarted);
(2)~application cloning (a new application is created and restarted
  from a previous checkpointed state of the original application); and
(3)~application migration (an application is cloned to another cloud and terminated 
on the source cloud).

In the first case, metadata for the checkpoints is retrieved from the Checkpoint Manager. 
Then the Application Manager triggers a {\em passive recovery mechanism}: new VMs 
can be restarted and provisioned if some VMs of the original set are not reachable any more.
Finally each VM in the computation downloads its corresponding checkpoint images from the storage. The process 
of restarting the application is delegated to DMTCP.

Cloning and migrating provide  alternative ways of creating an
application.  In these cases, a new application is created by issuing
a POST request to the {\tt coordinators resource}. Second, $n$~POST
requests are sent to the corresponding {\tt checkpoints resource}
to upload a set of checkpoint images. Finally, a POST to the {\tt
checkpoint resource} will restart the application. This will trigger
the {\em passive recovery mechanism} to generate a new virtual cluster
where the application will run.

\subsection{Application Termination}

Terminating an application consists of removing all references to
the application in the system. This can be decomposed as: (1)~deletion of the corresponding entry in the \emph{coordinator database};
(2)~removal of all the stored checkpoint images from the storage;
and (3)~release of the allocated VMs back to the pool of idle
VMs in the underlying infrastructure.

The TERMINATING state is reached when an end user issues a DELETE request
to the {\tt coordinator resource} or when the ERROR state is set for the application.

\section{Implementation of CACS}
\label{sec:prototype}
This section describes in detail some technical aspects of CACS.

\subsection{Cloud Manager}
The current  CACS prototype supports two underlying IaaS technologies
(Snooze and EC2 compatible VM management systems) allowing us to
demonstrate the portability and interoperability of CACS in
heterogeneous cloud environments.
 Snooze~\cite{Feller:2012:SSA:2310096.2310230} is a scalable  highly
available system. It has been primarily designed as a small system
easing the deployment of VMs and easing experimentation with VM management
strategies.  The Cloud Manager uses the specific REST API of Snooze
to manage virtual clusters. Snooze provides a server and VM failure
notification API that can be directly used by the Monitoring Manager.
(Thus with Snooze, no monitoring daemons need to be executed  in
the VMs.) EC2 compatible Cloud (like Openstack \cite{OpenStack})
are supported as well. OpenStack does not provide a failure
notification interface and thus the cloud-agnostic monitoring service
is used.

\subsection{Checkpoint Manager and Storage System}
The Checkpoint Manager, depicted in Figure \ref{fig:arch},
enables different storage systems to be plugged into CACS. The current
implementation of the service is stateless and  supports two storage 
systems: (1)~NFS and (2)~S3.  NFS is suitable for small-scale deployment
and especially for prototyping. S3 is the \emph{de~facto} standard API
of Amazon Web Service for manipulating stored objects. Supporting S3 gives CACS
compatibility with the major Cloud providers, but also with other
solutions such as Ceph. 
        
Since checkpoint images may be generated periodically, under
application control, or by the end user, a decision was made to
save checkpoint images asynchronously.  The Checkpoint Manager is
not aware of the existence of checkpoint images until a restart is
required.  At that time, the Checkpoint Manager will choose the
most recent checkpoint image, by default, but a user may also specify
an earlier image.

\subsection{Monitoring Manager}
\label{sec:failureDetection}
Some cloud platforms support an external API for monitoring if the
VMs are alive.  However, those cloud-specific mechanisms
are not sufficiently flexible.  Our goals are three-fold:
(1)~being cloud-agnostic; (2)~testing the liveness of the VMs; and 
(3)~testing the ``health'' of the application.

The concept of health is application-specific.
An application may fail due to unreachability of a computer node,
insufficient memory, internal busy waiting within an application, bugs
in the application code, issues induced by the execution context,
 the reception of spurious signals such as SIGCHILD and SIGPIPE,
and a myriad of other causes.  
A user-defined application-specific
routine can define and test the application's health using a function
hook offered by CACS.

The current implementation is based on a binary broadcast tree for
each application.  Each node of the broadcast tree is represented
by a daemon, which calls the user's hook function to determine if
the processes on that node are healthy.  A standard broadcast tree
then allows the root node to report a list of nodes that are unhealthy
or unreachable to the Monitoring Manager.  If problems are reported, 
the Monitoring Manager interacts with the Application Manager to trigger
an application recovery.

There are two cases:
\begin{enumerate}
  \item {\em VM failure:\/} A VM is unreachable.
CACS reserves a new VM from the underlying cloud infrastructure and
restarts the application from a previous checkpoint.
  \item {\em Application failure:\/}
If all VMs are reachable, the application itself may still be
reported as unhealthy.  As an optimization, one then kills the processes
of the application within their VMs, and restarts the
application processes within their {\em original} VMs.
\end{enumerate}

\subsection{Resilience: Avoiding Single Points of Failure}
CACS should be resilient to node failures. Its managers are stateless thus they can be easily restarted in the event of a failure. For purpose of high availability, traditional server replication and failover approaches leveraging  Apache Zookeeper~\cite{hunt2010zookeeper} can extend the current design. The coordinators database could be implemented relying on a NoSQL reliable distributed database technology such as Cassandra or MongoDB that does not exhibit any single point of failure.
The stable storage properties of the checkpoint storage are guaranteed through the use of a fault-tolerant distributed file system (e.g. Ceph) that provides persistent and highly available storage.

The Snooze IaaS cloud management system has been designed to be highly available in the event of simultaneous failures~\cite{Feller:2012:SSA:2310096.2310230}. Nevertheless, it does not ensure automatic recovery of virtual clusters in the event of the failure of the server hosting one of their VMs. By integrating CACS in Snooze,  computations running in virtual clusters can be automatically restarted in the event of a failure. Users of the enhanced Snooze system can enjoy both reliable application execution and a highly available IaaS cloud tolerating multiple simultaneous failures of physical machines hosting VMs and/or VM management services.

\subsection{Other Implementation Details}
CACS is implemented in Java and makes use of the scalable RESTlet~\cite{restlet} framework to expose
its API. The user requests are mostly treated in background using a pool of threads to optimize the parallelization and 
the responsiveness of the API. In the current implementation the coordinators database is stored in memory. The provision manager uses parallelization of SSH connection, to act on virtual clusters.

\section{Experimental Evaluation}
\label{sec:experimentalEvaluation}

\begin{figure*}[!ht]
  \centering
  \begin{subfigure}{0.3\textwidth}
    \includegraphics[width=\textwidth]{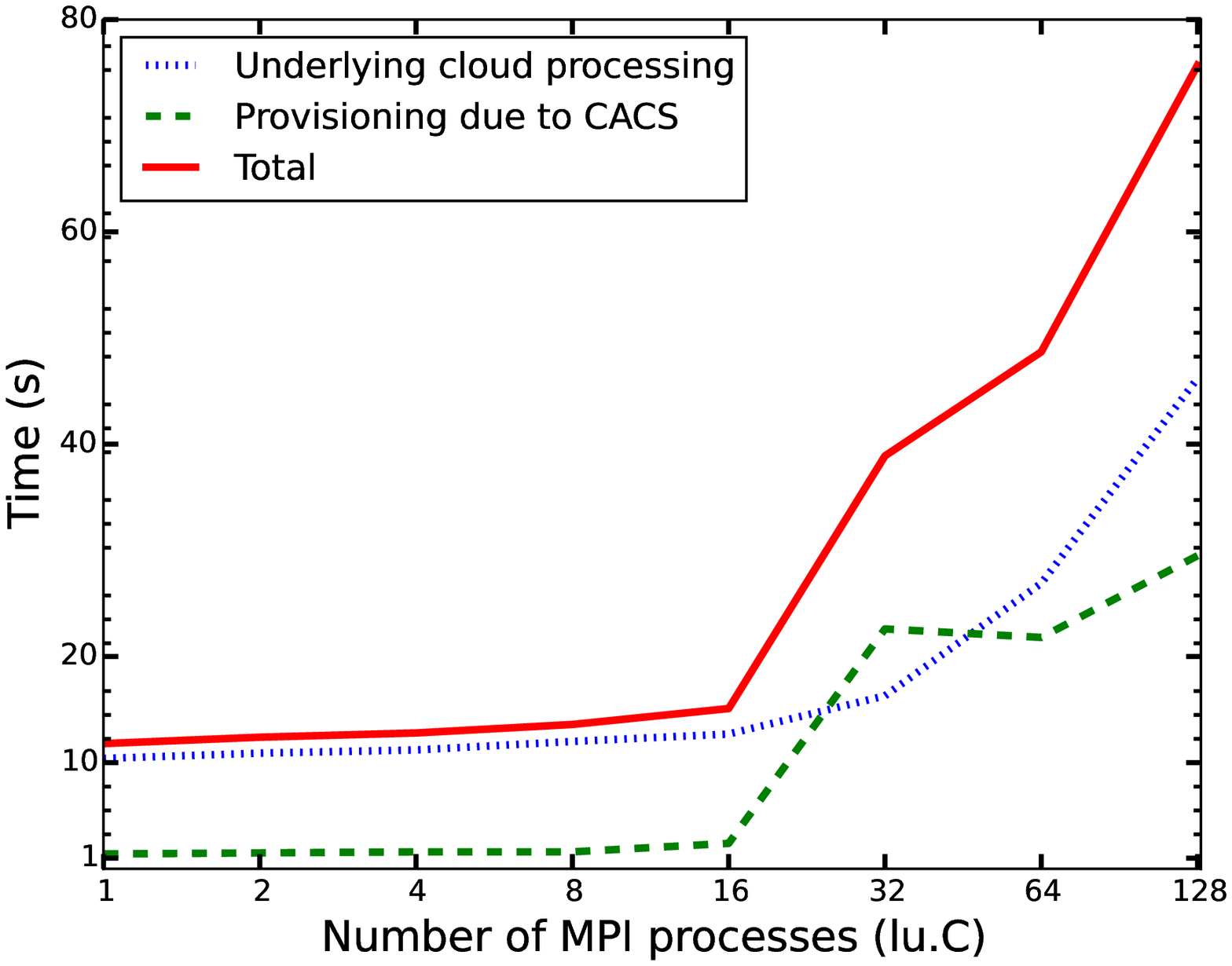}
    \caption{Submission time}
    \label{fig:snooze_submission}
  \end{subfigure}\hfill
  \begin{subfigure}{0.3\textwidth}
    \includegraphics[width=\textwidth]{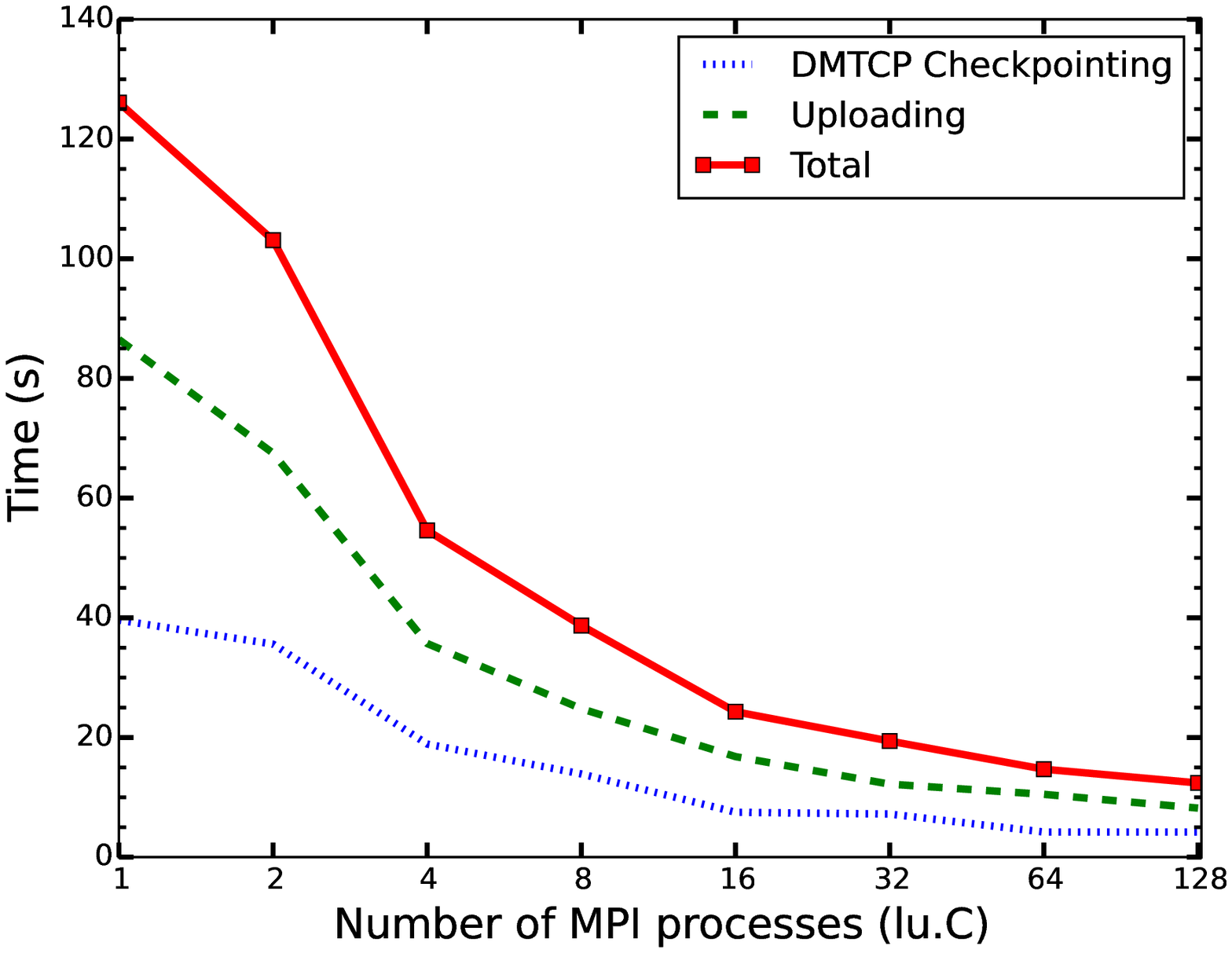}
    \caption{Checkpoint time}
    \label{fig:snooze_ckpt}
  \end{subfigure}\hfill
  \begin{subfigure}{0.3\textwidth}
    \includegraphics[width=\textwidth]{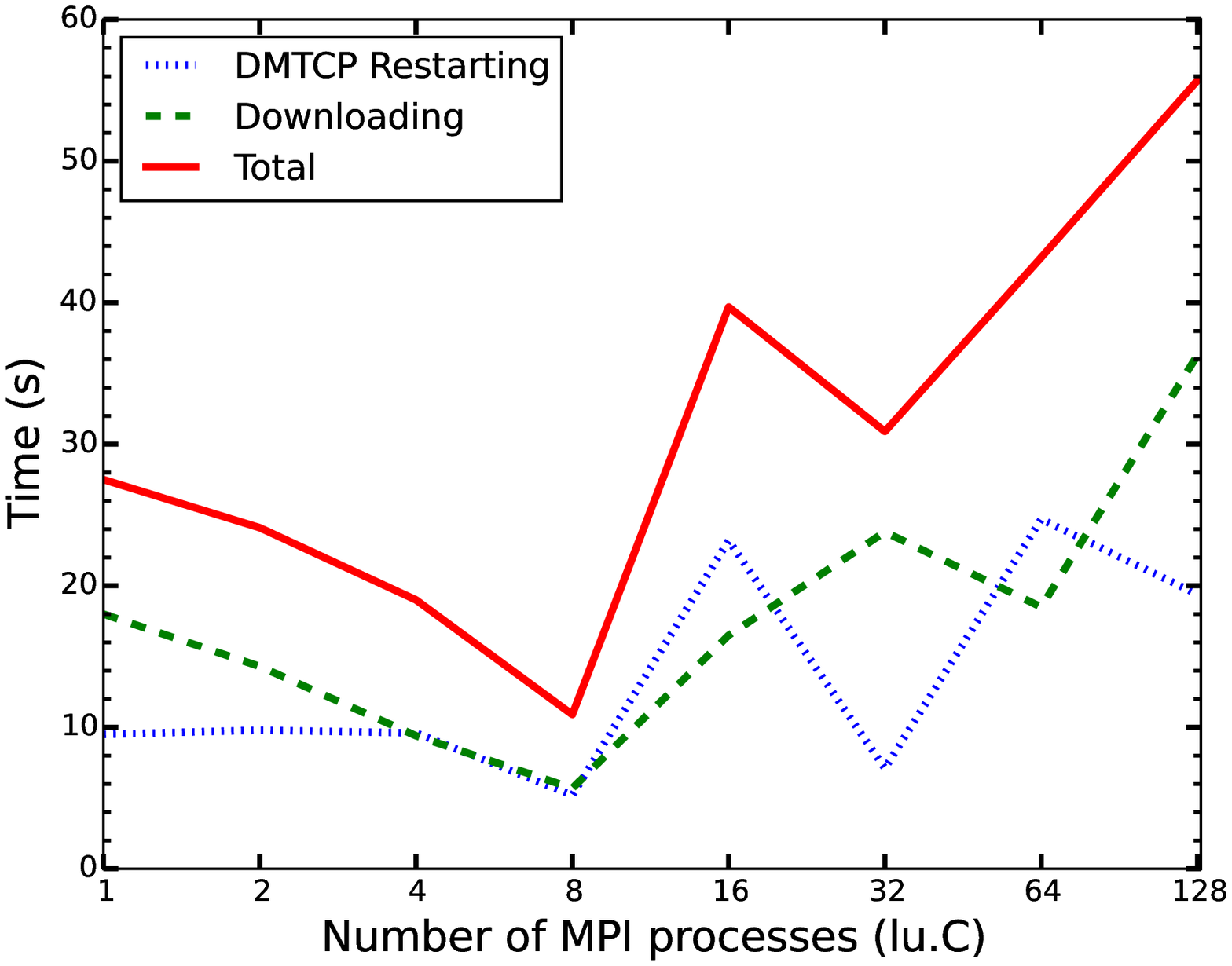}
    \caption{Restart time}
    \label{fig:snooze_restart}
  \end{subfigure}
  \caption{CACS over Snooze}
  \label{fig:scalability}
\end{figure*}

The experiments are divided into four parts: scalability with application size up to 128~nodes
using Snooze as a testbed (Section~\ref{sec:scalability}); resource
consumption of CACS, including the performance of the
monitoring system (Section~\ref{sec:consumption}); a performance study
of migration between different clouds or between desktop and cloud
(Section~\ref{sec:migration}); and a study of the cloud-agnostic feature
of CACS as applied to Snooze and OpenStack (Section~\ref{sec:comparison}).

The evaluation of the system was conducted on the Grid'5000~\cite{grid5000}
testbed.  A typical workflow for experimenting on
the platform is to reserve physical nodes, then to deploy a Linux-based
environment, and finally to deploy and configure the desired
software stack.  The Debian Wheezy distribution (3.2.0-4-amd64 Linux kernel)
served as the base
environment for deploying Snooze (version~2.1.6) and Ceph storage (Firefly).
Ubuntu 12.04 (kernel version 3.2.0-24-generic) was used for deploying Openstack (Icehouse). On the
two clouds we used an Ubuntu 13.10 x86\_64 base image, preconfigured
with the DMTCP distribution (version~2.3). Both clouds use KVM/QEMU.

\subsection{Scalability with Application Size}
\label{sec:scalability}

The scalability test was conducted using Snooze configured with more than 400~vCPUs
and nearly 1~TB of memory available, enough for holding more than 128
virtual machines, each of which requires one virtual core and 2~GB of memory. The NAS
MPI test for~LU (Class~C) was employed~\cite{bailey1991parallel}. Each MPI ran on a separate VM.
We measured the performance for three phases:
time to finish the application submission, time to perform a checkpoint,
and time to perform a restart. Figure~\ref{fig:scalability} shows that creation
of the VMs and execution of commands (provisioning, checkpoint,
restart) require significant time.  Time for submission depends strongly on the
underlying infrastructure used (see Section~\ref{sec:comparison} for more details),
while the latter two times are related to the number of VMs involved in the application. 

Figure~\ref{fig:snooze_submission} shows the performance for application submission,
which includes two steps: the underlying cloud allocates the VMs; and CACS
provisions the VMs. The proposed CACS
implementation optimizes the command execution mechanism through:
(1)~the parallelization of the SSH connections; and
(2)~re-use of the connections of the open SSH sessions. As a result, increasing the
number of nodes increases only slightly the time for executing commands, up until
the configured maximum limit of SSH connections is reached.
This occurs after 16~nodes in the current setup.

\begin{table}[!ht]
\centering
\begin{tabular}{|l|c|c|c|c|c|}
\hline
Number of processes   &   1   &   2   &   4   &   8   &   16   \\ \hline
Ckpt size (MB)        &   655 &   338 &   174 &   92  &   49   \\ \hline
\end{tabular}
\caption{\label{tbl:ckpt-size} Checkpoint image sizes for the NAS benchmark
lu.C, under different configurations. The checkpoint image size is for a
single MPI process.}
\label{tbl:ckpt_size}
\end{table}

The time for a single checkpoint is shown in
Figure~\ref{fig:snooze_ckpt}. Also, Table~\ref{tbl:ckpt_size} shows
the checkpoint image size as the number of nodes varies.  Here, the
primary workload contains two parts: DMTCP writes the checkpoint image
to local storage; and each VM uploads the image to the remote file
system. Figure~\ref{fig:snooze_restart} illustrates the performance for
restart. In this case, the trend becomes unstable for a large number of
nodes. This is due to network traffic when all nodes try to simultaneously
download the checkpoint images. As a consequence, restarted processes
do not join the computation concurrently, leading to jitter and less
reproducible timings for DMTCP restart.

\subsection{Resource Consumption and Monitoring System}
\label{sec:consumption}

This section focuses on the resource consumption of CACS,
as well as the performance of the monitoring system. They share the same
experimental configuration: for Snooze, 7~servers hosting VM management services and 12~servers hosting VMs (264~cores in total) were deployed. The target application used
was {\em dmtcp1}, a single-process lightweight application found in the DMTCP
test suite~\cite{AnselEtAl09}. 

\begin{figure*}[!ht]
  \centering
  \begin{subfigure}{0.3\textwidth}
    \includegraphics[width=\textwidth]{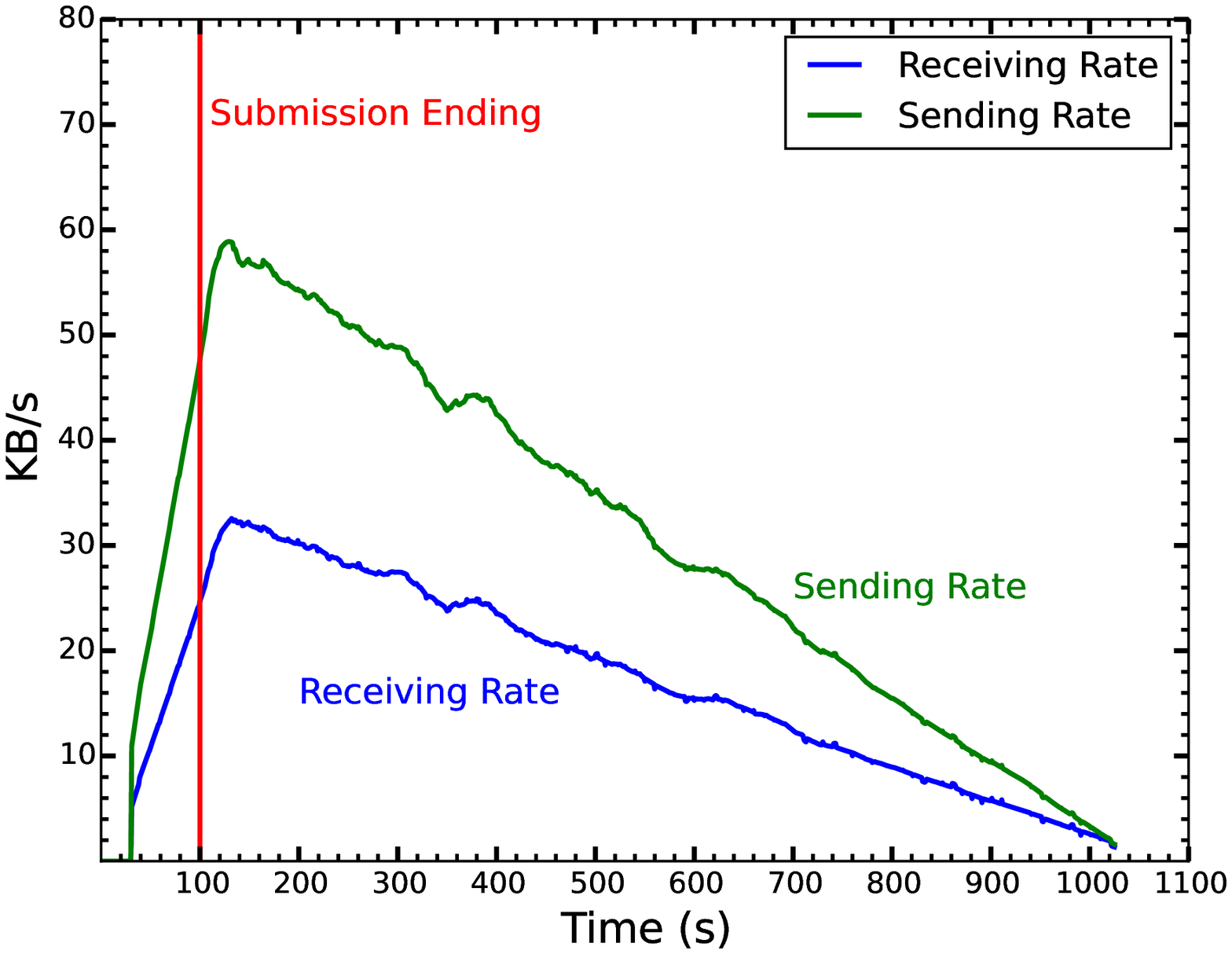}
    \caption{Network bandwith \\ ~}
    \label{fig:network_consumption}
  \end{subfigure}\hfill
  \begin{subfigure}{0.3\textwidth}
    \includegraphics[width=\textwidth]{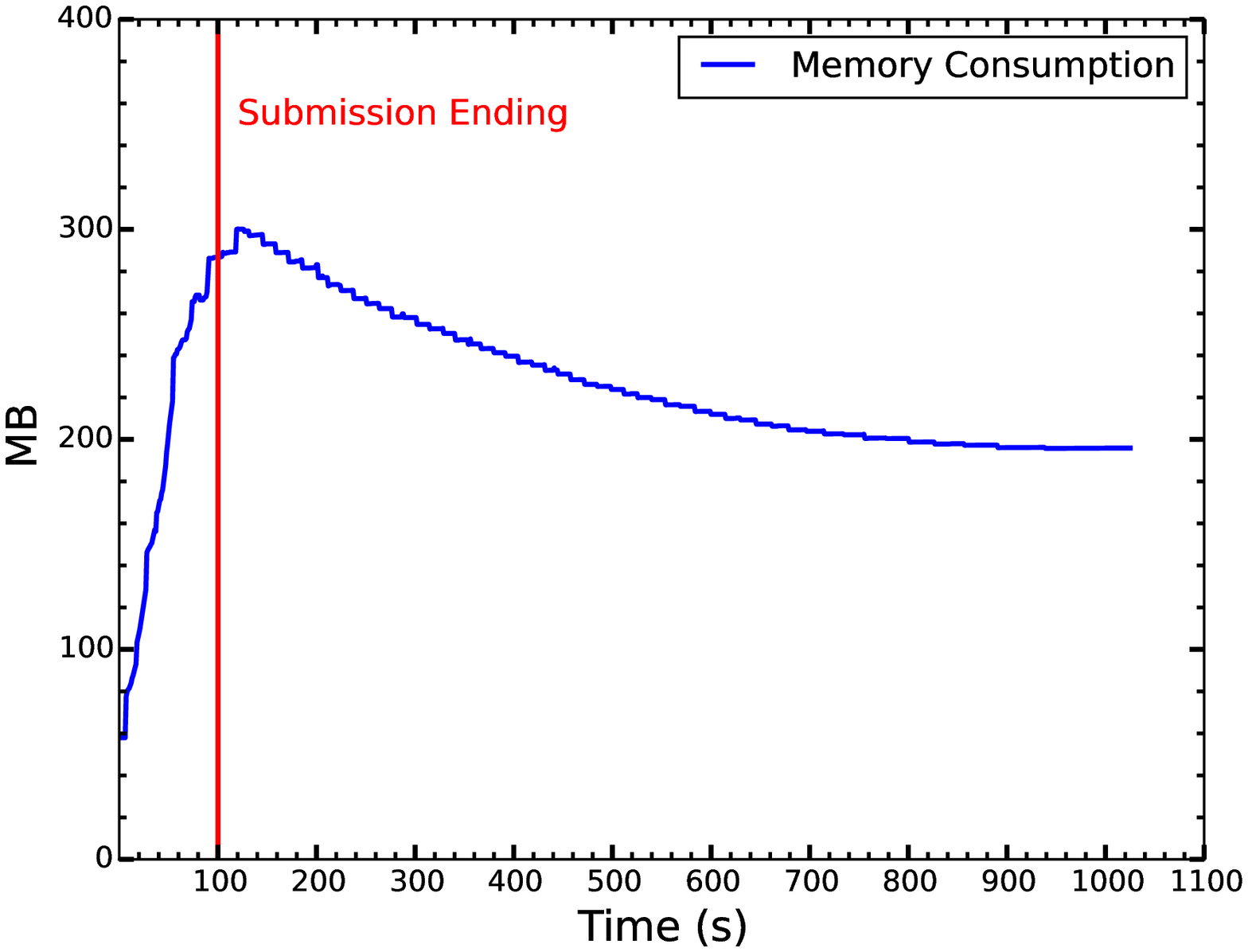}
    \caption{Memory consumption \\ ~}
    \label{fig:memory_consumption}
  \end{subfigure}\hfill
  \begin{subfigure}{0.3\textwidth}
    \includegraphics[width=\textwidth]{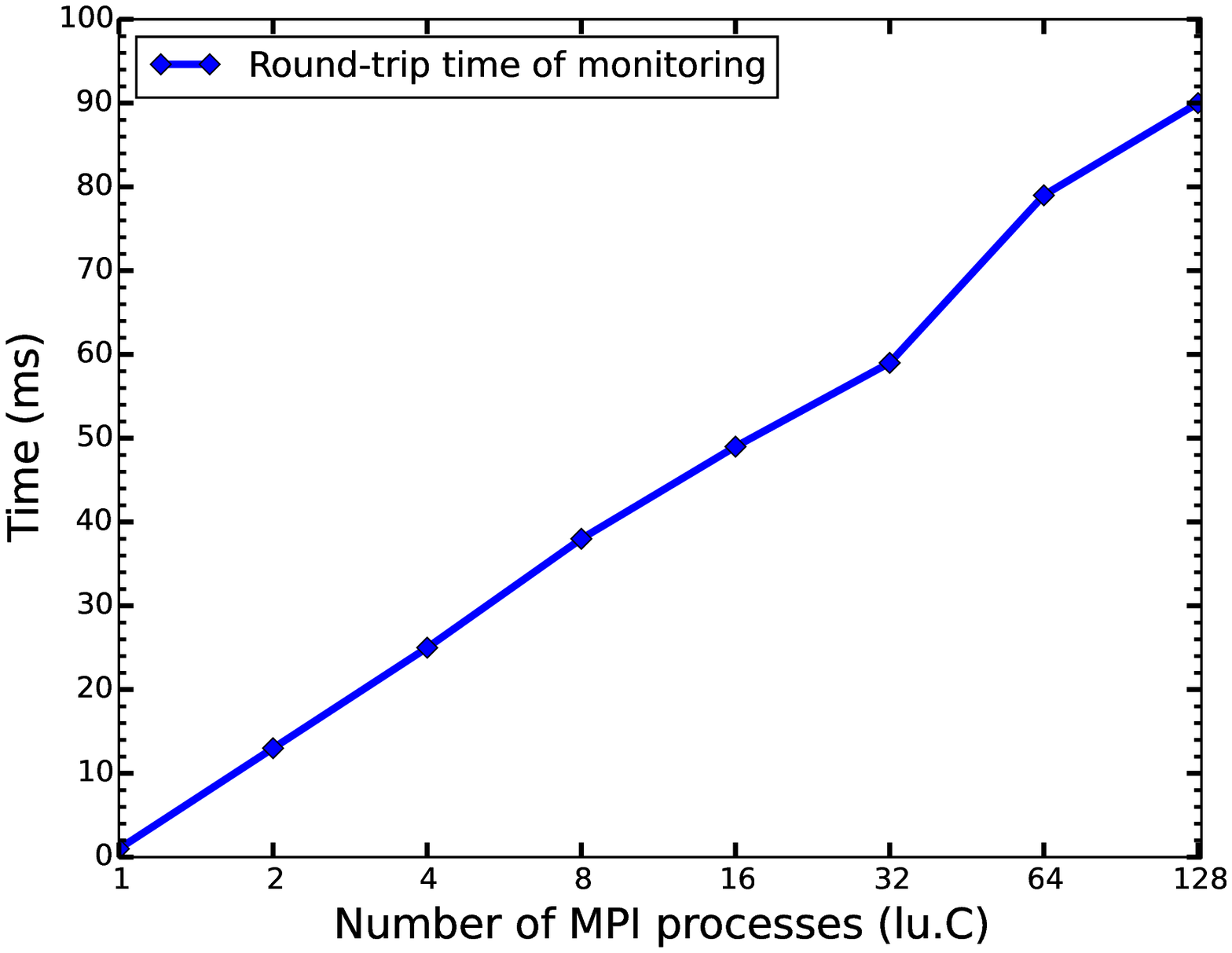}
    \caption{Performance of monitoring system
             (Note: logarithmic x-axis)}
    \label{fig:monitoring_performance}
  \end{subfigure}
  \caption{CACS resource evaluation}
  \label{fig:resource_consumption}
\end{figure*}

\subsubsection{Resource Consumption of the Service}

In this experiment,~100 applications were submitted to CACS,
with one new application submitted every second.
The network consumption and memory usage
are shown in Figure~\ref{fig:network_consumption} and
Figure~\ref{fig:memory_consumption}, respectively. The vertical line
at~100 seconds shows when the 100~applications had been submitted, but
were not necessarily executing yet. Both
figures show a decreasing trend as processing continues.

Figures~\ref{fig:network_consumption} and~\ref{fig:memory_consumption}
can be understood better through a review of the CACS implementation.
CACS maintains a thread pool to handle incoming submissions.
Theoretically, it can concurrently handle as many applications as
there are threads in the thread pool (100~in this experiment).  But
the underlying cloud infrastructure has its own limitations: most
clouds can handle only a relatively small number of applications
concurrently.

The linear decline in network bandwidth observed after the vertical
line in Figure~\ref{fig:network_consumption} can be explained as
follows.  Assume that the cloud can handle~$n$ submissions concurrently,
implying that there are~$n$ threads running SSH commands on the VMs
provided by the cloud.  Meanwhile, there are~$m$ threads polling
the cloud front-end as it causes the VMs to be built. Assume also
that the network bandwith consumed by a polling thread and an SSH
thread are both constants, namely,~$c\_1$ and~$c\_2$. Based on these
assumptions, we conclude that at any given time, the network traffic
is:
\begin{align*}
  &m c\_1 + n c\_2.
\end{align*}
In our case, $m$ is initially~100.  Since VMs are
processed at a uniform rate, $m$~will decrease at a uniform rate,
thus explaining the linear trend in Figure~\ref{fig:network_consumption}.
A similar analysis also explains the decreasing trend seen in
Figure~\ref{fig:memory_consumption}.

\subsubsection{Performance of the Health Monitoring System}

The health monitoring system was discussed in
Section~\ref{sec:failureDetection}.
To measure its performance, we
submitted applications with varying numbers of VM requests, and
recorded the time required to finish one round-trip for a heartbeat
(employing the binary broadcast tree described earlier).
Figure~\ref{fig:monitoring_performance} shows the result: the time
to finish one heartbeat round-trip is logarithmic in the number
of nodes, as expected.  This provides strong evidence that
the broadcast tree implementation consumes few
network resources and scales to support large distributed applications.

\subsection{Migration Evaluation}
\label{sec:migration}

Migration of distributed applications are important in the real
world.  CACS is evaluated in two migration scenarios.
Section~\ref{sec:cloudification} evaluates the {\em cloudification}
of an NS-3~\cite{NS-3} application. NS-3 simulations are known for requiring
long periods of time, and thus are good candidates for migration from
commodity hardware to the cloud.  Section~\ref{sec:cloud_migration}
demonstrates the migration of applications between two distinct clouds:
Snooze and OpenStack.

\subsubsection{From Hardware to Cloud}
\label{sec:cloudification}

{\em Cloudification} refers to migrating a conventional desktop or laptop
application to the cloud.  Statistics were obtained for migrating an
NS-3 application from a physical machine to the OpenStack destination cloud.
The target application was {\em tcp-large-transfer} from the NS-3
distribution.  The parameters of the application were set to simulate a
1~Gb~transfer rate transferring 2~GB of data over a period of 30~seconds.
The application was checkpointed after 10~seconds.  A 50-line Python
script invokes CACS, which checkpoints on the current machine
and restarts in the destination cloud. The application contains a single
process and the checkpoint image was approximately 260 MB.
Application restart on OpenStack required 21~seconds. Note that in
the destination cloud none of the VMs have NS-3 installed.  DMTCP
checkpoint images include a copy of the memory of the process.
Since the NS-3 libraries were already present in memory, they were
transported to the destination cloud as part of the checkpoint
images.

\subsubsection{From Cloud to Cloud}
\label{sec:cloud_migration}
Next, application migration between Snooze and OpenStack was studied. Two
instances, CACS-Snooze and CACS-Openstack, were deployed each relying on
its corresponding IaaS platform. The target application is dmtcp1,
the same as in Section~\ref{sec:consumption}.  40~different instances of
the application were incrementally started on CACS-Snooze and then  cloned
to CACS-OpenStack using a 90-line Python script.  The script
checkpoints on the current cloud and restarts on the destination
cloud.  The experimental setup used a single instance of Ceph-based
storage for both services, since both clouds were deployed on
Grid'5000.  Alternatively, two distinct storage systems could have
been used as well with no modification. The application checkpoint
periodicity was set to~60 seconds.  The checkpoint image sizes were
approximately 3~MB each.

\begin{figure}[h!]
  \centering
    \includegraphics[width=0.5\columnwidth]{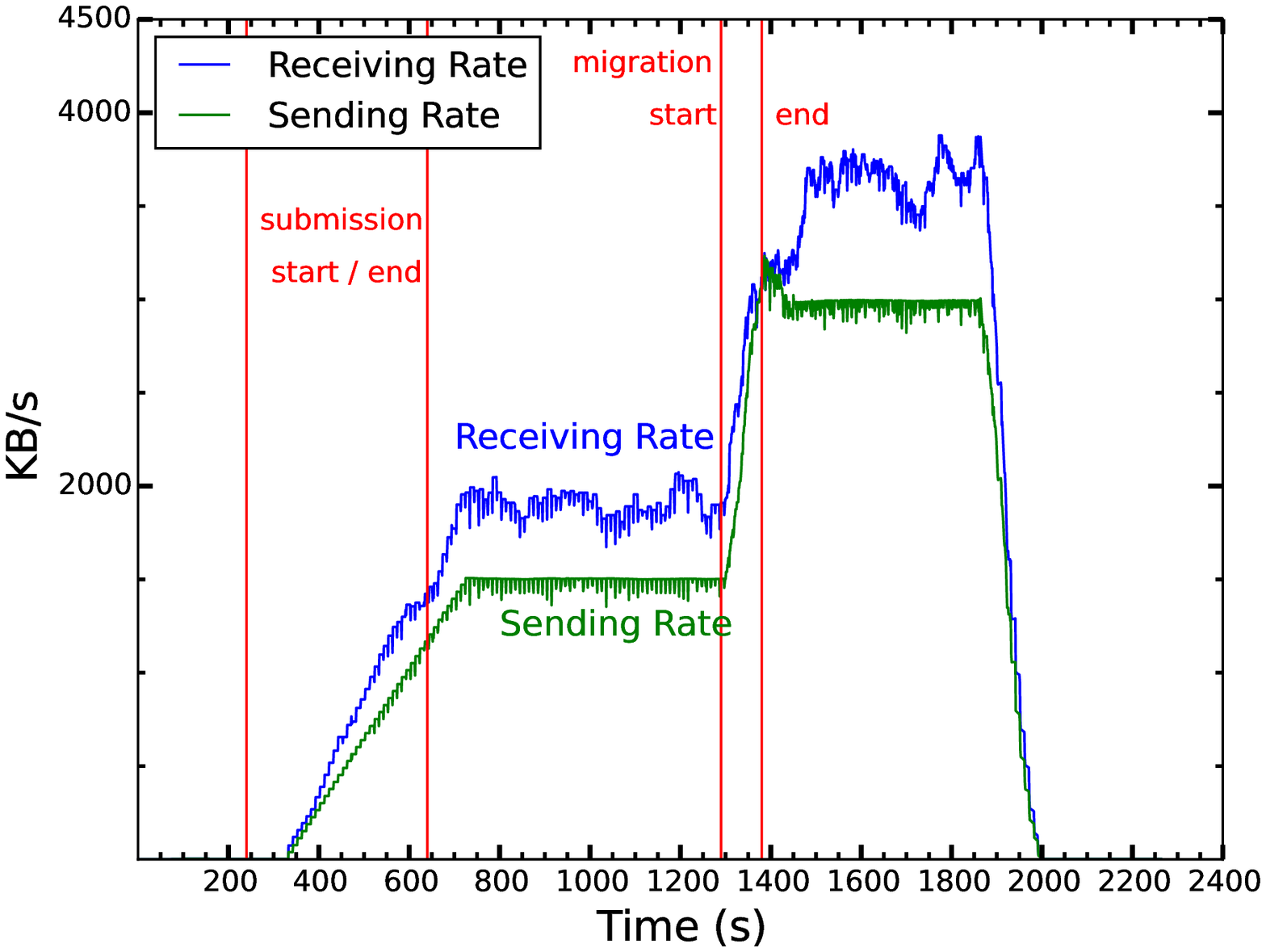}
  \caption{Migration performance of 40 applications}
  \label{fig:migration_40}
\end{figure}

Figure~\ref{fig:migration_40} depicts the overall network utilization
at the storage level. It shows a linear increase of network utilization
after start of the submissions. A plateau indicates that the applications'
checkpoint images were received and stored and no submissions remain. The
migration phase lasts for 2.5~minutes.  The network utilization
during this phase increases due to the data transfer of the checkpoint
images. Note that the time to transfer checkpoint images from CACS-Snooze
to CACS-OpenStack is negligible in this case, due to the small size of
the checkpoint images. The network utilization then reaches another plateau,
indicating that two instances of each application are now running on the
two different clouds (80 applications in total). After a certain period of time, all applications
are terminated.

The experiment also demonstrates the ability of CACS to handle numerous concurrent restart requests (up to 40  requests). 

\subsection{Comparison of Different IaaS Technologies}
\label{sec:comparison}
\begin{figure}[h!]
  \begin{subfigure}{0.5\textwidth}
  \centering
    \includegraphics[width=0.9\columnwidth]{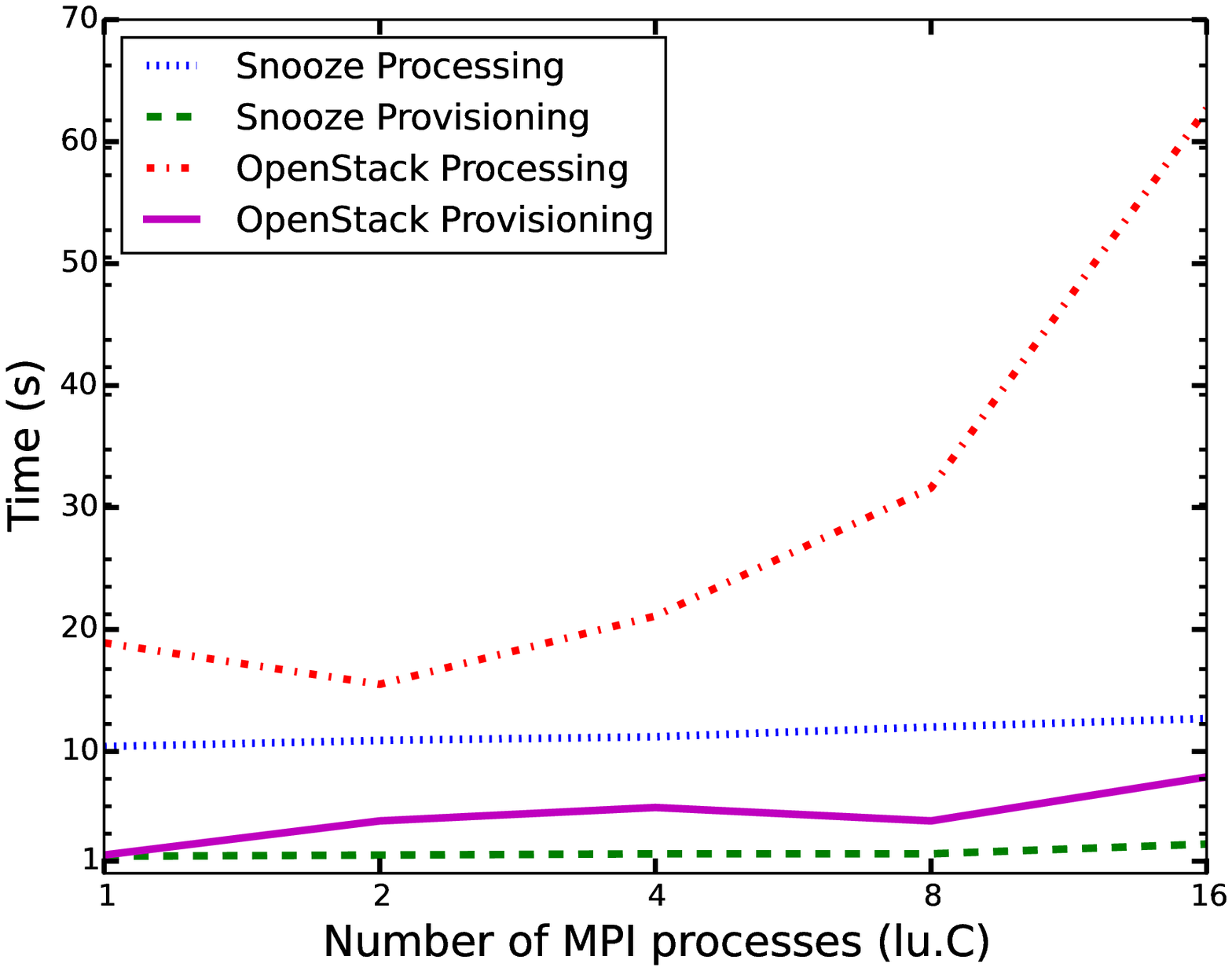}
  \caption{Comparison of submission time}
  \label{fig:comparison_submission}
  \end{subfigure}
  \begin{subfigure}{0.5\textwidth}
  \centering
    \includegraphics[width=0.9\columnwidth]{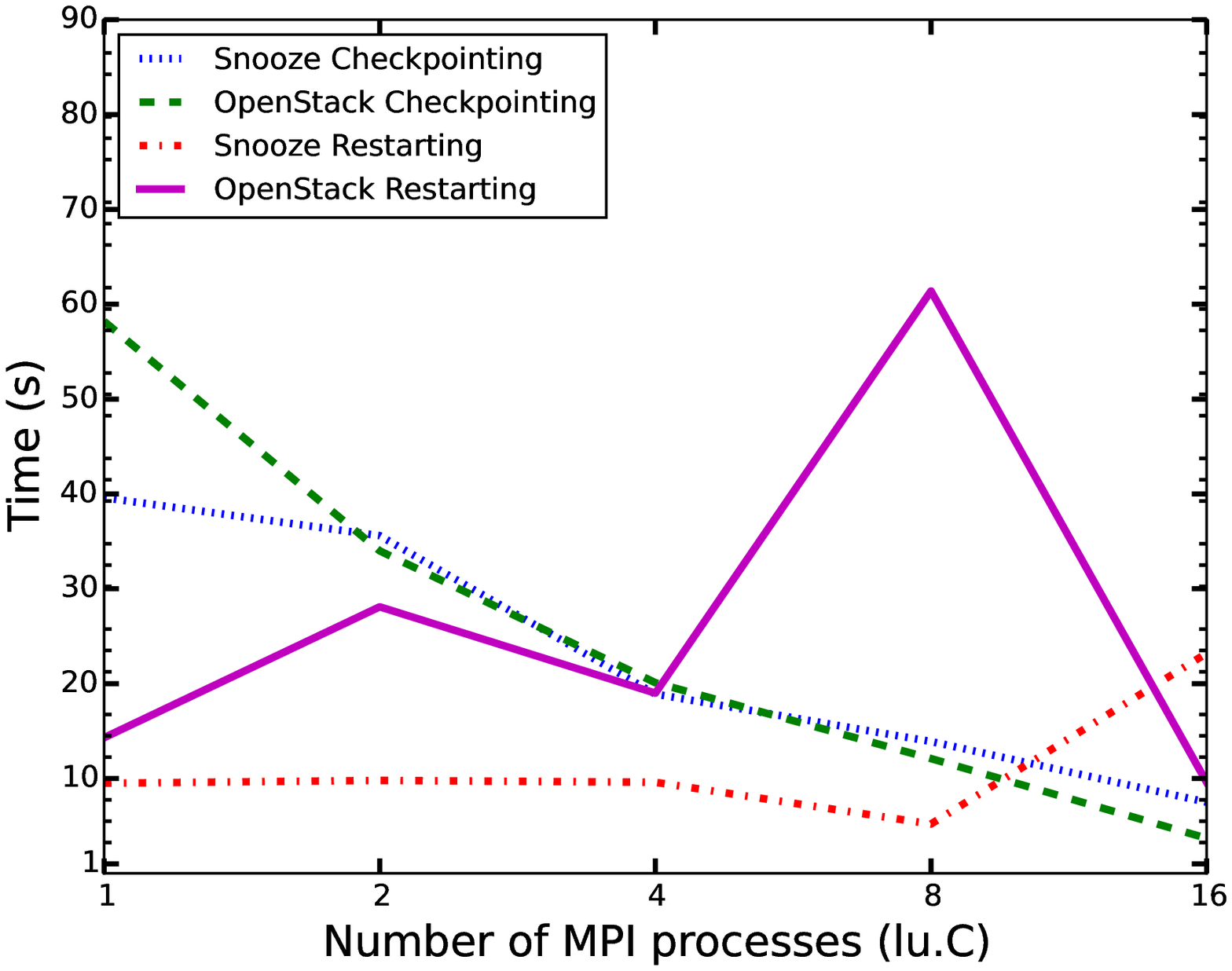}
  \caption{Comparison of Checkpoint/Restart time}
  \label{fig:comparison_ckpt_restart}
  \end{subfigure}
  \caption{Comparison of CACS for Snooze and OpenStack}
\end{figure}

Next, we compare the performance of CACS, when targeted
toward two distinct cloud management
systems: Snooze and OpenStack. The configuration for Snooze is the
same as in Section~\ref{sec:scalability}, while OpenStack is configured
with the same computing resources.
Figure~\ref{fig:comparison_submission} reports the
submission times, including both the time for the IaaS to process the VM
submissions, and the time for CACS to provision the~VMs.
Figure~\ref{fig:comparison_ckpt_restart} reports
the checkpoint and restart times.  Note that the same
checkpoint policy was used for both clouds.  Hence, the checkpoint sizes are the
same (see Table~\ref{tbl:ckpt_size}).  This implies that
the uploading time during checkpoint and the downloading time
during restart should be comparable, except to the extent that
different network traffic conditions exist during the two phases.
For this reason, Figure~\ref{fig:comparison_ckpt_restart} reports a
single time for checkpoint/restart, since the times for checkpoint and
restart are comparable.

Figure~\ref{fig:comparison_submission} shows that although
the underlying IaaS systems are different, the time for CACS 
to provision the VMs remains comparable.  In contrast, the time for
different IaaS systems to process VM allocation differs greatly. The
preceding breakdown into CACS-specific and IaaS-specific times illustrates
that CACS is able to support different cloud management
systems, with little or no CACS-specific difference in performance.

Figure~\ref{fig:comparison_ckpt_restart} shows a similar trend, except
that the restart time for OpenStack is not stable. This occurs for
a pragmatic reason: normally, OpenStack recommends that network
management and applications be located on different networks.
However, the limitations of the Grid'5000 testbed forced the placement of
both types of network data onto the same network.  This leads to
data variability, as seen in the figure.

\section{Related Work}
\label{sec:relatedWork}

We first review previous work on application checkpointing and virtual machine
snapshot mechanisms. We then study existing approaches for reliable application
execution in clouds.

\subsection{Options for Checkpointing Distributed Applications}
\label{sec:distribCheckpoint}

In addition to DMTCP, several other checkpointing packages are
in common use today.  The survey~\cite{Egwutuoha2013} describes several
checkpoint/restart implementations for high performance computing.
More generally, we review the checkpoint-restart packages in widespread
use today.

For distributed computation, most checkpointing services today were
built around MPI-specific checkpoint-restart services. 
 Unfortunately, this was not an option for the
current work, since a cloud-agnostic checkpointing service must also
be application-agnostic.  Nevertheless, historically there has been
much effort toward MPI-specific custom checkpoint-restart service.
This came about when InfiniBand became the preferred network for high
performance computing, and there was still no package for transparent
checkpointing over InfiniBand.  Examples of checkpoint-restart services
can be found in Open~MPI~\cite{Hursey07} (CRCP coordination protocol),
LAM/MPI~\cite{Sankaran05} (now incorporated into MVAPICH2~\cite{Gao06}),
and MPICH-V~\cite{Bouteiller06}.  Each checkpoint-restart service
disconnected the network prior to checkpoint, called on a single-node
checkpointing package such as the kernel-based BLCR~\cite{BLCR06},
and then re-connected after restart.

The current work is based on a new approach, implemented within DMTCP,
which enables transparent checkpointing over InfiniBand~\cite{CaoEtAl14}
as well as over~TCP.  This uniformly supports both MPI and other distributed
applications.

\subsection{Mechanisms Based on VM Snapshots}

VM snapshots provide an alternative for
checkpointing.  The well-known packages KVM/QEMU, VirtualBox, and
VMware all support snapshots of VMs.  However, the choice
of a particular VM for a particular cloud platform is
contrary to the goal of a cloud-agnostic service in this work.
Furthermore, saving just the application is more efficient than
saving an entire VM, in part due to the smaller
memory footprint of a bare application.

\subsection{Fault Tolerance and Efficiency in the Cloud}
There exist several alternative approaches to fault tolerance in the Cloud.
Tchana \hbox{et al.} argue for shared responsibility between provider
and user~\cite{TchanaEtAl12}.
Zhao \hbox{et al.} follow a middleware approach~\cite{ZhaoEtAl10}.
Egwutuoha \hbox{et al.} take a process-level redundancy
approach~\cite{EgwutuohaEtAl12}.
Di \hbox{et al.} present an adaptive algorithm to optimize the number
of checkpoints for cloud-based applications~\cite{DiEtAl13}.

Nicolae \hbox{et al.} show how to checkpoint an MPI application
using distributed VM snapshots using the BlobSeer
distributed repository~\cite{NicolaeEtAl11}.  This approach
is MPI-specific, since it employs the MPI checkpoint-restart
service with BLCR. (See Section~\ref{sec:distribCheckpoint} for a
discussion of MPI checkpoint-restart services.)  Kangarlou \hbox{et
al.}~\cite{kangarlou2012vnsnap} and Garg \hbox{et al.}~\cite{GargEtAl13}
each show how to take a distributed snapshot of VMs.
Kangarlou \hbox{et al.} base this on a modification of Xen's live
migration, while Garg \hbox{et al.} employ DMTCP to take a distributed
snapshot of KVM/QEMU VMs.  The last three investigations
(\cite{NicolaeEtAl11}, \cite{kangarlou2012vnsnap}, \cite{GargEtAl13})
contrasts with the cloud-agnostic (and application-agnostic) approach
employed here by directly checkpointing the processes along with their network
connections.

Several works study detection of failure
modes~\cite{SchroederEtAl10,XiongEtAl12}. The approach of the Gamose system~\cite{gamose} for monitoring the
health of Grid applications can extend the CACS checks for application
health without requiring application hooks.  Such a system relies on
interposing on systems calls, and does not require any modification to
the operating system.

The work of Marshall \hbox{et al.}~\cite{5948611} demonstrates
the utility of backfill VMs in maintaining a high utilization
rate for the processors of the cloud.  The backfill VMs
can be combined with a checkpointing policy so as to always guarantee
a supply of checkpoint images that can be restarted on demand to instantiate
the backfill VMs.

\section{Conclusion}
\label{sec:conclusion}

The Cloud-Agnostic Checkpointing Service
demonstrates checkpointing
as a service on top of heterogeneous IaaS cloud systems in an environment
of multiple heterogeneous clouds. A key design principle of CACS is
that it is built around the DMTCP mechanism for taking checkpoints of
distributed processes, rather than employing distributed VM snapshot
mechanisms.  This creates a cloud-agnostic service that is independent
of the cloud platform, and independent of the cloud's choice of virtual
machine technology. Preliminary experimental evaluations demonstrate
portability between two IaaS cloud management systems
and demonstrate scalability with the number of applications
and the application size.  CACS also supports migration
between heterogeneous clouds, and {\em cloudification}, migration from
a traditional environment to the cloud.
 In our  next steps, we will further improve the scalability  of CACS, 
 study its efficiency in different computing environments varying the resource, VM and 
 storage management  systems, and experiment with a broader range of distributed applications.

\section*{Acknowledgment}

Experiments presented in this paper were carried out using the
Grid'5000 experimental testbed, being developed under the INRIA
ALADDIN development action with support from CNRS, RENATER and
several universities as well as other funding bodies (see
https://www.grid5000.fr).

\bibliographystyle{alpha}
\bibliography{dmtcp-snooze}

\end{document}